\documentclass[aps,prc,twocolumn,showpacs,subfigure,superscriptaddress,nobibnotes,nofootinbib]{revtex4-1}
\usepackage[dvips]{graphicx}
\usepackage{epsfig}
\usepackage{bm}   
\usepackage{dcolumn}
\usepackage{graphicx}
\usepackage{rotate}
\usepackage{amsfonts}

\begin{document}

\def\vlk{$V_{{\rm low}\, k}$}
\def\cor{$^{88}$Sr}
\def\zr {$^{92}$Zr}
\def\zra{$^{94}$Zr}
\def\nb {$^{93}$Nb}
\def\mo {$^{94}$Mo}
\def\ru {$^{96}$Ru}
\def\pd {$^{98}$Pd}
\def\cd {$^{100}$Cd}
\def\gs {$0^+_1$}
\def\ms {2$^+_{\rm I,ms}$}
\def\s  {2$^+_{\rm I   }$}
\def\ss {2$^+_{\rm II  }$}
\def\mss{2$^+_{\rm II,ms}$}
\def\sc{1$^+_{\rm sc}$}

\def\veff{$V_{eff}$}
\def\vnn{$V_{NN}$}
\def\etal{\it et al.}
\def\pn{\it pn}
\def\pp{$\pi\left(p_{1/2}\right)$}
\def\pg{$\pi\left(g_{9/2}\right)$}
\def\nda{$\nu\left(d_{5/2}\right)$}
\def\ndb{$\nu\left(d_{3/2}\right)$}
\def\ns{$\nu\left(s_{1/2}\right)$}
\def\ng{$\nu\left(g_{7/2}\right)$}
\def\nh{$\nu\left(h_{11/2}\right)$}
\def\nbgs{$\frac{1}{2}^-_1$}
\def\nbsa{$3/2^-_1$}
\def\nbsb{$5/2^-_1$}
\def\nbmsa{$3/2^-_2$}
\def\nbmsb{$5/2^-_2$}
\def\nbsc{$3/2^-_3$}
\def\nbsd{$5/2^-_3$}
\def\fm {fm$^{-1}$}
\def\mun{$\mu_N^2$}
\def\prl{Phys.\ Rev.\ Lett.\ }
\def\pra{Phys.\ Rev.\ A }
\def\prb{Phys.\ Rev.\ B }
\def\prc{Phys.\ Rev.\ C }
\def\prd{Phys.\ Rev.\ D }
\def\np {Nucl.\ Phys.\ }
\def\pl {Phys.\ Lett.\ }
\def\rmp{Rev.\ Mod.\ Phys.\ }
\def\pr {Phys.\ Rep.\ }

\title{Fermionic-bosonic couplings in a weakly-deformed odd-mass nucleus, $^{93}_{41}$Nb
\\}

\author{J. N.~Orce}
\email{jnorce@triumf.ca} \homepage{http://www.pa.uky.edu/~jnorce}
\affiliation{Department of Physics and Astronomy, University of
 Kentucky, Lexington,
 Kentucky 40506-0055, USA}
\affiliation{TRIUMF, 4004 Wesbrook Mall, Vancouver, BC V6T 2A3, Canada}

\author{J. D. Holt}
\affiliation{TRIUMF, 4004 Wesbrook Mall, Vancouver, BC V6T 2A3, Canada}
\affiliation{Physics Division, Oak Ridge National Laboratory, P.O. Box 2008,
Oak Ridge, TN 37831, USA}
\affiliation{Department of Physics and Astronomy, University of Tennessee,
Knoxville, TN 37996, USA}

\author{A. Linnemann}
\affiliation{Institut f\"ur Kernphysik, Universit\"at zu K\"oln,
50937 K\"oln, Germany}

\author{C. J. McKay}
\affiliation{Department of Physics and Astronomy, University of
Kentucky, Lexington, Kentucky 40506-0055, USA}

\author{C. Fransen}
\affiliation{Institut f\"ur Kernphysik, Universit\"at zu K\"oln, 50937 K\"oln, Germany}

\author{J. Jolie}
\affiliation{Institut f\"ur Kernphysik, Universit\"at zu K\"oln,
50937 K\"oln, Germany}

\author{T.T.S. Kuo}
\affiliation{Nuclear Structure Laboratory, Department of Physics and
Astronomy, SUNY, Stony Brook, NY 11794-3800, USA}

\author{\\S. R. Lesher}
\affiliation{Department of Physics and Astronomy, University of
Kentucky, Lexington, Kentucky 40506-0055, USA}
\affiliation{Department of Physics, University of Wisconsin - La Crosse, 1725 State Street, La Crosse, WI 54601, USA}

\author{M. T. McEllistrem}
\affiliation{Department of Physics and  Astronomy, University of Kentucky, Lexington,
Kentucky 40506-0055, USA}

\author{N. Pietralla}
\affiliation{Institut f\"ur Kernphysik, Universit\"at zu K\"oln,
50937 K\"oln, Germany}
\affiliation{Institut f\"ur Kernphysik, Technische Universit\"at Darmstadt, D-64289 Darmstadt, Germany}

\author{N. Warr}
\affiliation{Institut f\"ur Kernphysik, Universit\"at zu K\"oln, 50937 K\"oln, Germany}

\author{V. Werner}
\affiliation{Institut f\"ur Kernphysik, Universit\"at zu K\"oln,
50937 K\"oln, Germany}
\affiliation{Wright Nuclear Structure Laboratory, Yale University, New Haven, CT 06520-8120, USA}

\author{S. W. Yates}
\affiliation{Department of Physics and Astronomy, University of Kentucky, Lexington,
Kentucky 40506-0055, USA}
\affiliation{Department of Chemistry, University of Kentucky, Lexington, Kentucky
40506-0055, USA}

\date{\today}

%
%

\begin{abstract}

A comprehensive decay scheme of $^{93}$Nb below 2 MeV has been
constructed from information obtained with the
$^{93}$Nb(n,n$^\prime$$\gamma$)
 and  $^{94}$Zr(p,2n$\gamma$$\gamma$)$^{93}$Nb reactions. Branching ratios, lifetimes,
transition multipolarities and spin assignments have been determined. From $M1$ and
$E2$ strengths, fermionic-bosonic excitations of isoscalar and
isovector character have been identified from the weak coupling $\pi
1g_{9/2}$$\otimes$$^{92}_{40}$Zr and $\pi
2p_{1/2}^{-1}$$\otimes$$^{94}_{42}$Mo configurations.
A microscopic interpretation of such excitations is attained from
 shell-model calculations using low-momentum
effective interactions.

\end{abstract}


\maketitle
%


\section{Introduction}

Low-lying collective excitations in weakly deformed, odd-mass
nuclei have rarely been studied in detail due to the structural complexity
arising from the interplay of various single-particle excitations  
and collective degrees of freedom. With increasing angular momentum  for
the ground and single-particle states, bountiful
excitations and a distribution of strength among several levels occurs; 
however, the weak coupling of a single particle to the bosonic core 
would provide structural simplifications. 
Such a scenario is present in the
low-energy structure of the weakly deformed odd-mass nucleus, $^{\ 93}_{\ 41}$Nb$_{52}$, which 
exhibits unique, 
unmixed structures built on the $\pi 1g_{9/2}$ ground state and the $\pi
2p_{1/2}^{-1}$ proton-hole excited state. 

The magnetic dipole operator, $\hat M1$, can be decomposed into an 
isoscalar ($IS$) and an isovector ($IV$) term, 
\begin{eqnarray}
\hat M1 &=& IS~(\Delta T =0) + IV~(\Delta T =\pm 1) \\  \nonumber
&&\propto   \sum_{i=1}^{N,Z}  ~\mu_N\left[ ~\frac{1}{2}~(g_{\pi}^s + g_{\nu}^s)~\boldsymbol{s} + ~(g_{\pi}^l + g_{\nu}^l)~\boldsymbol{\ell} ~\right]   
\\  \nonumber
&&+ \sum_{i=1}^{N,Z} ~\mu_N\left[~\frac{1}{2}~ (g_{\pi}^s - g_{\nu}^s)~\boldsymbol{s} ~+  ~(g_{\pi}^l - g_{\nu}^l)~\boldsymbol{\ell} ~\right] 
\boldsymbol{\tau_3} , 
\label{m1}
\end{eqnarray}
\noindent where $\boldsymbol{\ell}, \boldsymbol{s}$ and  $\boldsymbol{\tau_3}$ are the orbital, spin and z-component isospin operators,  
respectively, $\mu_N$  the nuclear magneton and $g_\pi^l=1\mu_N$, $g_l^n=0$ and $g_s^p=5.586\mu_N$, $g_s^n=-3.826\mu_N$ are the orbital 
and spin proton and neutron $g~factors$ in free nuclear matter, respectively. 
The $IS$ term 
is much weaker than the $IV$ one due to the strong cancellation of 
the spin $g~factors$.  Isovector excitations  occur when nucleons and
their spins are collectively excited.   Proton 
and neutron spin {g factors} are additive in the vector part of
the magnetic dipole ($M1$) operator, which  may lead to large $M1$ 
transition strengths.   The scissors mode~\cite{5} in deformed nuclei~\cite{sm1,nord,nord2,sm2} 
and mixed-symmetry ($MS$) states in weakly-deformed even-even nuclei~\cite{msreport} are examples of collective 
isovector excitations. 
The latter  was predicted  by the
algebraic Interacting Boson Model-2 (IBM-2) with separate
representations for proton and neutron bosons~\cite{1}, and interpret as a  
collective motion of proton and neutrons not in phase. Such
isovector excitations have widely been identified in
weakly-deformed even-even nuclei~\cite{msreport}. Both 
$IS$ and $IV$ excitations have  been identified in the negative-parity
structure of $^{93}$Nb built on the 2$p_{1/2}^{-1}$ proton-hole excited
state~\cite{orce}. Given the large quadrupole moment, $Q_s(9/2^+_1)=-0.32(2)$, 
recently determined for the ground state~\cite{93nb_qmoment},  the invocation of a simpler vibrational picture seems
unnecessary, and this work provides a
microscopic many-body interpretation of such isovector excitations.

In fact, $MS$ states have  been explained from a shell-model ($SM$)
basis by Lisetskiy and collaborators~\cite{3} using a surface 
delta interaction with tuned parameters. Large isoscalar $E2$ matrix elements were found between
states with $MS$ assignments and interpreted as excitations with
similar $pn$ symmetry. In a microscopic many-body framework, shell-model calculations using
a $^{88}$Sr core and effective low-momentum interactions {\vlk}~\cite{bogner03} have
successfully accounted for excitation energies, level densities,
$M1$ and $E2$ transition strengths, isoscalar and isovector
excitations, and both spin and orbital contributions to the $M1$
transition matrix elements in the negative-parity structure of
$^{93}$Nb~\cite{orce}. As a successor to that work, we present a more
extensive spectroscopic and microscopic study of $^{93}$Nb,
including both positive- and negative-parity structures.


\vspace{-0.5cm}
\section{Experimental details}

The nucleus $^{93}$Nb has been studied with the (n,n$'$$\gamma$)
reaction at the University of Kentucky and in a 
$^{94}$Zr(p,2n)$^{93}$Nb $\gamma$-$\gamma$ coincidence experiment at
the University of Cologne.

\subsection{$^{93}$Nb(n,n$'$$\gamma$) experiments}

Excitation functions, lifetimes and branching ratios were measured
using the $^{93}$Nb(n,n$^\prime$$\gamma$) reaction \cite{nng} at
neutron energies ranging from 1.5 to 2.6 MeV. Neutrons were provided
by the 7-MV electrostatic accelerator at the University of Kentucky
through the $^3$H(p,n)$^3$He reaction. The scattering sample was 56 g 
of Nb, which is naturally monoisotopic, in a 3$\times$2 cm cylinder. 
Pulsed-beam techniques were used to reduce background,
with beam pulses separated by 533 ns and bunched to about 1 ns.
The time-of-flight technique for background
suppression was employed by gating on the appropriate prompt time windows
\cite{nng}. Finally, $\gamma$ rays were detected using
a BGO Compton-suppressed 55\% HPGe spectrometer with 2.0 keV
resolution. Both excitation functions and angular distributions were
normalized to the neutron flux.


From the angular distribution experiments, lifetimes were determined
through the Doppler-shift attenuation method following the
(n,n$'$$\gamma$) reaction \cite{belgya}. Here, the shifted
$\gamma$-ray energy is given by,

\vspace{-0.3cm}
\begin{equation}
E_{\gamma}(\theta_\gamma)=E_{\gamma_{0}}
[1+\frac{v_{0}}{c}F(\tau)cos{\theta_\gamma}],
\end{equation}

\noindent with $E_{\gamma_{0}}$ being the unshifted  $\gamma$-ray
energy, $v_{0}$ the initial recoil velocity in the center of mass
frame, $\theta$ the angle of observation and $F(\tau)$ the
attenuation factor, which is related to the nuclear stopping
process described by Blaugrund \cite{blaugrund}. Finally, the
lifetimes of the states can be determined by comparison with the
$F(\tau)$ values calculated using the Winterbon formalism
\cite{winterbon}.

Furthermore, $\gamma$-ray coincidences were observed through the
(n,n$'$$\gamma$$\gamma$) reaction at a neutron energy of 3 MeV in
order to identify the decay paths of $\gamma$ rays, measure their
branching ratios, and confirm the results from the excitation functions. 
The coincidence methods
have been described by McGrath {\it et al.}~\cite{mcgrath}.

\begin{figure}[]
\begin{center}
\includegraphics[width=7.cm,height=7.6cm,angle=-0]{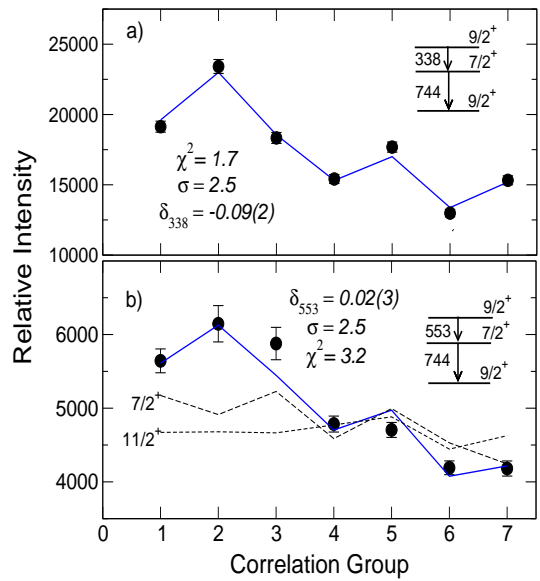}
\caption{(Color online) Angular correlation plots for a) the 338-keV and b) the 553-keV 
transitions gated by the 744 keV $\gamma$ ray. The fits to the data give  mixing ratios of
$\delta_{338}=-0.09(2)$ (in agreement with
previous work, $\delta_{338}=-0.12(5)$)  and a more precise $\delta_{553}=0.02(3)$ as compared with 
previous work, $\delta=-0.3^{+3}_{-7}$~\cite{nndc}. 
Dashed lines indicate other
angular correlation fits with unlikely spins for the  1297.4 keV
level. The fitting parameter $\sigma$ is the width of the magnetic
substate distribution, usually  found to be between  2 and 3. }
\label{fig:1}
\end{center}
\end{figure}


\subsection{$^{94}$Zr(p,2n)$^{93}$Nb angular correlation experiment}

For  odd-mass nuclei, particularly those with high
ground-state spins, the angular distributions of the 
photons from scattering reactions are rather isotropic, and  the (n,n$^{\prime}
\gamma$) data cannot be used generally for assigning the spins of excited states. Complementarily,
from a $\gamma$-$\gamma$ coincidence experiment with the
$^{94}$Zr(p,2n$\gamma\gamma$)$^{93}$Nb reaction, many multipolarities and spin
assignments could be determined, and branching ratios were measured. The
proton beam provided by the 10-MV tandem accelerator at the
University of Cologne bombarded a 2 mg/cm$^2$ thick $^{94}$Zr
target enriched to 96.93$\%$ with beam currents from 2 to
3 $\mu$A. The proton energies were in the range of 11.5 to 19 MeV. The de-exciting
$\gamma$ rays were detected using the HORUS spectrometer~\cite{horus}, comprised
of 16 HPGe detectors with the following features: a) seven of the detectors 
formed a EUROBALL cluster detector~\cite{euroball} with the
central detector placed at $\theta$=90$^{\circ}$ with respect to the
beam axis and $\phi$=0$^{\circ}$, where $\phi$ is the clockwise
angle around the beam axis; b) four detectors placed at
$\theta$=45$^{\circ}$ and 135$^{\circ}$ in a vertical plane above
and below the beam axis were complemented by Compton suppression
shields; c) the remaining detectors  were placed in the
$\theta$=90$^{\circ}$ plane and at angles of $\phi$=55$^{\circ}$,
125$^{\circ}$, 235$^{\circ}$ and 305$^{\circ}$.
When possible, we determined the $E2/M1$ mixing ratio, $\delta^2=\Gamma_{f,E2}/\Gamma_{f,M1}$ \cite{delta},
by fitting the angular correlation data as a function of the angles, $\theta_1$, $\theta_2$ and $\phi$,
following the formalism
developed by  Krane and Steffen \cite{angcorr}.
The current results are  in  agreement with the most intense 
transitions identified in previous work
\cite{93nb_old,kakavand}. As  an example of this agreement, the top
panel of Fig. \ref{fig:1} shows  an angular  correlation plot for
the previously known 338-keV $\gamma$ ray depopulating
the 1082-keV one-phonon excitation. A value of
$\delta_{744}=0.26(8)$  was fixed from the angular correlation
analysis of $\gamma$ rays populating the 744-keV level.

\section{Data analysis and results}

In previous work by van Heerden and McMurray \cite{93nb_old},
angular momenta, parities and transition rates for several of the
first ten states (up to 1.1 MeV) were presented. The
low-lying structure of $^{93}$Nb is dominated by the
$\pi 1g_{9/2}$ and $\pi 2p_{1/2}^{-1}$ single-particle and proton-hole excitations, 
respectively, and $jj$
couplings built on these 
\cite{ins1,ins2,transfer,kakavand,93nb_old}. In particular, the
$\pi 1g_{9/2}$ ground state couples to the 2$^+_1$ state of
the $^{92}$Zr core to give a quintet of  isoscalar excitations, J$^{\pi}$ $=$ 5/2$^+$,
7/2$^+$, 9/2$^+$, 11/2$^+$ and 13/2$^+$. The 
3/2$^-$ and 5/2$^-$ states are also  identified as being built on the
low-lying $\pi 2p_{1/2}^{-1}$ excitation. Our results are in 
agreement with previous work.

\begin{figure}[!]
\begin{center}
\includegraphics[width=6.cm,height=8.cm,angle=-90]{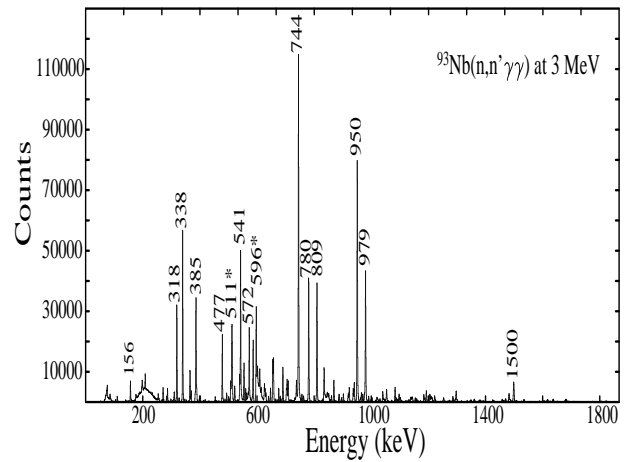}
\caption{Background-subtracted total projection of the
$\gamma$-$\gamma$-coincidence matrix from  the
$^{93}$Nb(n,n$'$$\gamma$$\gamma$) reaction at an incident neutron energy of 3
MeV. The labelled transitions belong to $^{93}$Nb, except for those indicated by asterisks (electron-positron
annihilation and radiation produced by neutrons striking the HPGe detectors.)} \label{fig:tot}
\end{center}
\end{figure}

\begin{figure}[!]
\begin{center}
\hspace{-0.1mm}\includegraphics[width=5.cm,height=8.cm,angle=-90]{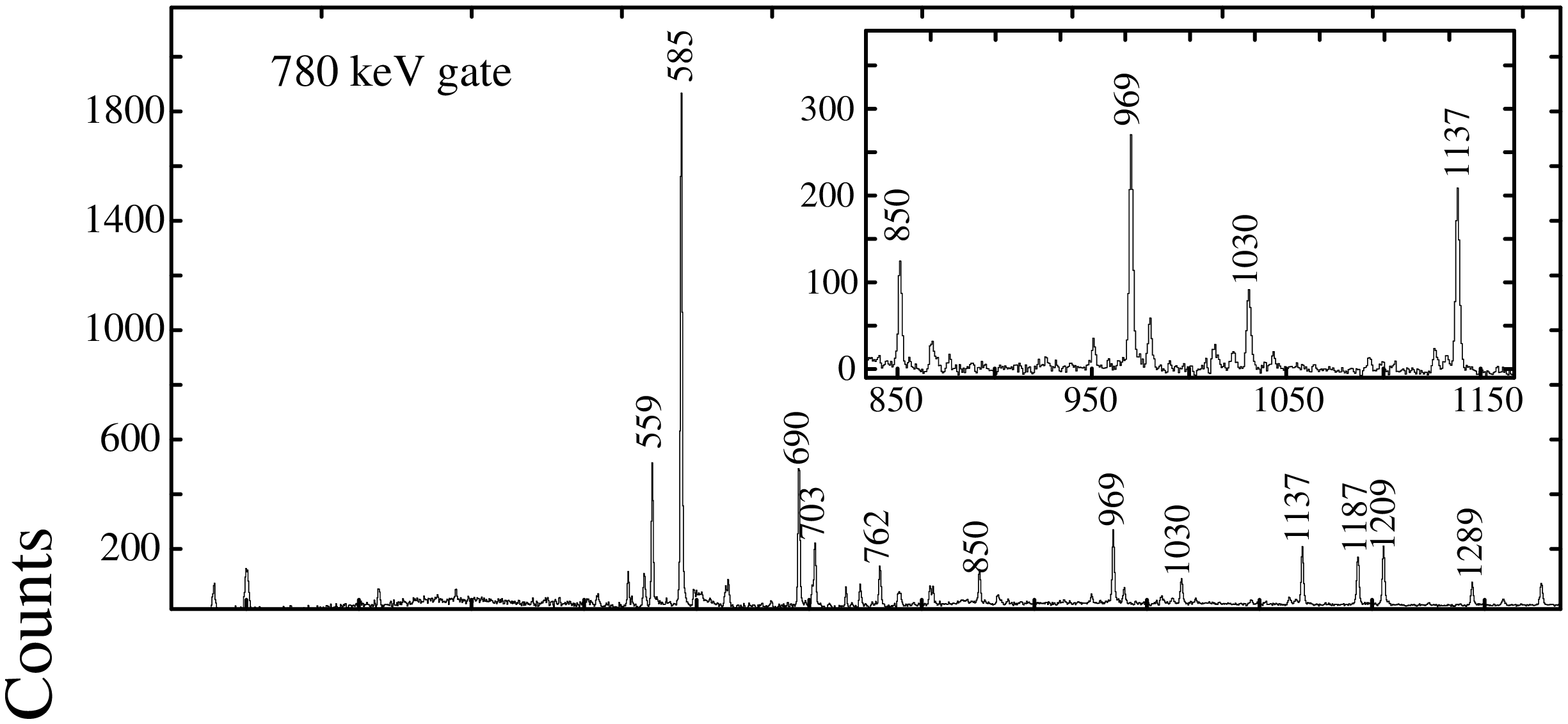}\\
\vspace{-2.99mm}\includegraphics[width=5.cm,height=8.cm,angle=-90]{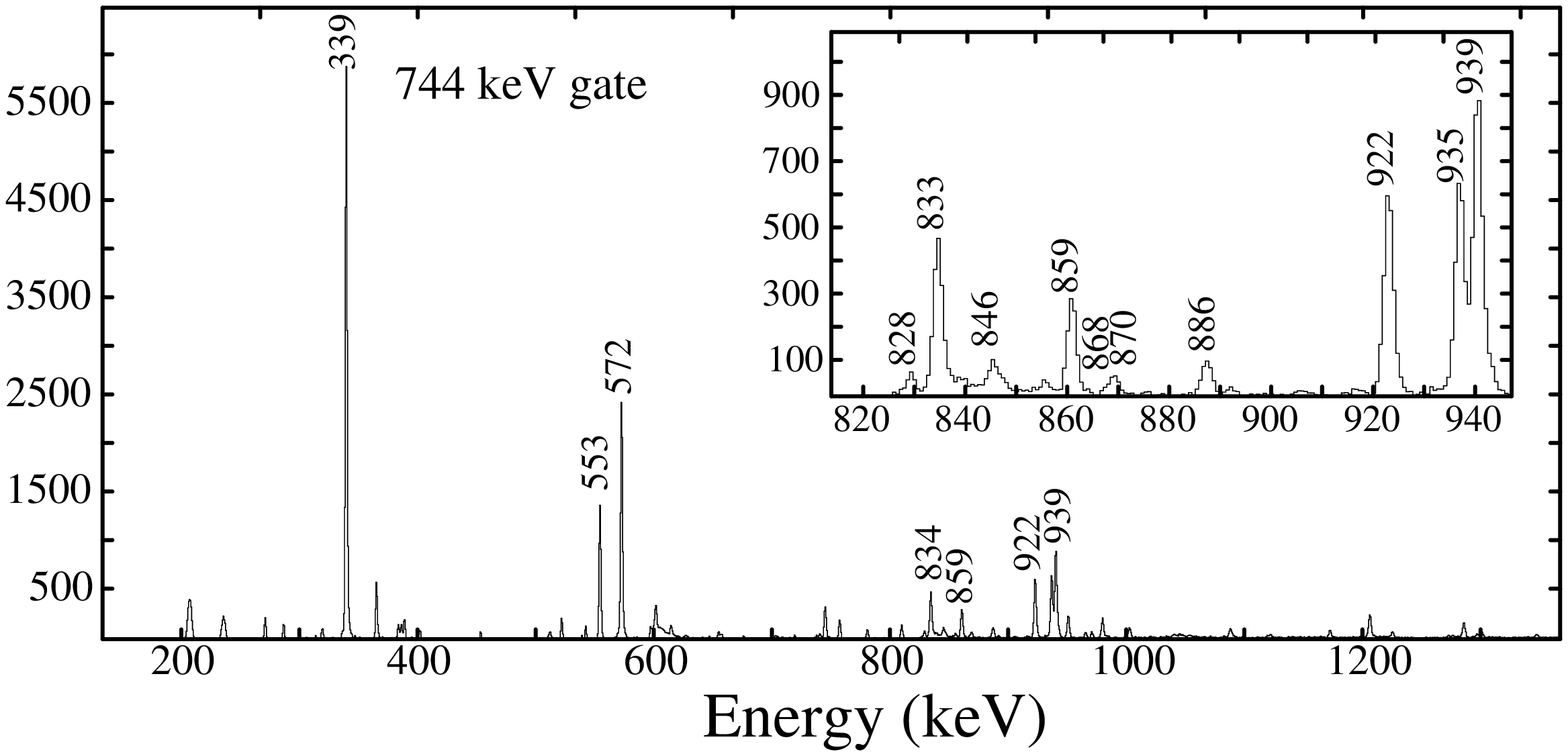}
\caption{Background-subtracted $\gamma$-ray spectra gated by the
744- and the 780-keV transitions in the $^{93}$Nb(n,n$'$$\gamma$$\gamma$)
coincidence matrix.} \label{fig:744gg}
\end{center}
\end{figure}

Excitation functions, together with the analysis of gated
coincidence spectra,  allowed the construction of a comprehensive level
scheme up to 2.2 MeV. Figure~\ref{fig:tot} shows the total projection
of the coincidence matrix built from the
$^{93}$Nb(n,n$'$$\gamma$$\gamma$) experiment, with the main
$\gamma$-ray transitions depopulating the nucleus labeled. The data also support 
two well-defined and nearly separate structures
as the main characteristic of this nucleus.  Figure~\ref{fig:744gg} shows typical
background-subtracted spectra gated on the 744-keV and 780-keV
transitions feeding the ground state and 1/2$^-$ state,
respectively, indicating the different decay patterns as well as the
quality of the data. These two separate structures are presented and
discussed according to the parity of the states and their
$IS/IV$ character.

\begin{table*}[]
\begin{center}
\caption{Properties of low-lying positive-parity states in $^{93}$Nb
below 2.2 MeV. Excitation energies,  spins, lifetimes,
$\gamma$-ray energies, branching and mixing ratios, initial and
final spins of the states, and reduced
transition probabilities are listed. An asterisk labels newly identified
levels and $\gamma$-ray transitions. Uncertainties in the energies
are 0.2 keV.} \label{tab:pos}
\renewcommand{\baselinestretch}{0.5}\small\small
\begin{tabular}{lccccccccccc}
\hline \hline \\
 E$_{level}$       &    $\tau$           &  $J^{\pi}_i$  $\rightarrow$  $J^{\pi}_f $ & $E_{\gamma}$     & $I_{\gamma}$ &  $\delta$    &  $B(M1)$    &  $B(E2)$  &  \hspace{3mm} $B(M1)$ & $B(E2)$  \\
 (keV)             &     (fs)            &                                           & $(keV)$          &              &              & ($\mu_N^2)$ &  (W.u.)   &  \multicolumn{2}{c}{\hspace{3mm}[Shell Model]} &&   \\  \hline \\
           0.0     &    stable           &                                           &                  &              &              &          &        &    &       \\ \\
        744.0     &                     &  7/2$^{+}$    \hspace{5mm}  9/2$^{+}$     &   744.0          &    100       &     0.28(8)  &     &    & 0.09591   & 7.836 \\ \\  
         809.1     &                     &  5/2$^{+}$    \hspace{5mm}  7/2$^{+}$     &    65.0          &    $\ <$1    &     (M1/E2)  &          &        & 0.1023  & 3.80      \\
                   &                     &               \hspace{13mm}  9/2$^{+}$    &   808.6          &    100(2)    &      E2      &          &        & -  & 11.313\\ \\
         949.8     &                     &  13/2$^{+}$   \hspace{5mm}  9/2$^{+}$     &   949.8          &    100       &      E2      &          &        & -  & 8.050  \\ \\

         979.6     & 715$^{+350}_{-180}$ &  11/2$^{+}$   \hspace{5mm}  9/2$^{+}$     &   978.8          &    100       &   $-$0.13(7) &          &        & 0.04328  & 5.761 \\ \\

        1082.6     &    $\ >$1245        &   9/2$^{+}$   \hspace{5mm}  11/2$^{+}$    &   103.7$^*$      &    3(2)      &     M1/E2    &             &            & 1.747  & 1.091 \\
                   &                     &               \hspace{13mm}  7/2$^{+}$    &   338.6          &  100(2)      &     M1/E2    &             &            & 0.1795  & 2.299  \\
                   &                     &               \hspace{13mm}  9/2$^{+}$    &  1082.6          &   38(2)      &     M1/E2    &             &            & 0.00731  & 3.992 \\ \\

   1127.0$^{IS}$  &                     &   7/2$^{+}$   \hspace{5mm}   5/2$^{+}$    &   318.3          &   100(2)     & $-$0.20(6)   &             &            & 0.016  & 0.6033\\ \\

   1297.4$^{IV}$   & 380$^{+110}_{-75}$  &   9/2$^{+}$   \hspace{5mm}  11/2$^{+}$    &   318.3          &    31(5)     &   (M1/E2)    &  $\ <$0.9   &            & 0.02185  & 1.420 \\
                   &                     &               \hspace{13mm}  7/2$^{+}$    &   553.1          &    61(5)     & $-$0.03(5)   &   0.33(5)   &   0.6(1)   & 0.2476  & 2.299  \\
                   &                     &               \hspace{13mm}  9/2$^{+}$    &  1297.4          &   100(5)     &  0.31(9)     &   0.04(1)   &   1.2(2)   & 0.00366  & 0.0591 \\ \\

   1315.7$^{IV}$  & 530$^{+450}_{-170}$  &   5/2$^{+}$   \hspace{5mm}    5/2$^{+}$   &   506.7          &    24(4)     &  $-$1.4(8)   &    0.05(4)  &  235$^{+300}_{-200}$  &0.190  &0.3816  \\
                  &                      &               \hspace{13mm}   7/2$^{+}$   &   571.5          &   100(4)     &   0.14(4)    &    0.45(25) &  15(10)     & 0.4072  & 3.864 \\ \\

   1334.9$^{IS}$  &                      & 17/2$^{+}$   \hspace{5mm}    13/2$^{+}$  &    385.1          &   100(2)    &      E2      &             &            &        &  \\ \\

   1483.6$^{IV}$   &     65(5)           &   7/2$^{+}$   \hspace{5mm}     9/2$^{+}$  &   400.8$^*$      &    7(2)      &   (M1/E2)    &  $\ <$0.7   &            & 0.3459 & 0.02121 \\
                   &                     &               \hspace{13mm}    5/2$^{+}$  &   674.8          &   27(2)      &  $-$0.11(8)  &   0.58(6)   &   9(1)     & 0.1970 & 0.1264 \\
                   &                     &               \hspace{13mm}    9/2$^{+}$  &  1483.8          &  100(2)      &  $-$0.13(7)  &   0.20(2)   &   0.9(6)   & 0.00422 & 5.568  \\ \\

   1490.9$^{IS}$  &   $\ >$750          &  15/2$^{+}$   \hspace{4mm}     17/2$^{+}$ &   156.0          &   14(2)      &   (M1/E2)    &   $\ <$0.04 &            &  &  \\
                   &                     &               \hspace{14mm}    13/2$^{+}$ &   541.1          &  100(2)      &  $-$0.11(2)  &   $\ <$0.47 &   $\ <$12  &  &  \\ \\

   1603.5$^{IS}$  & 465$^{+250}_{-125}$ &  11/2$^{+}$    \hspace{4mm}     9/2$^{+}$   &    520.9          &   13(3)      &  $-$0.07(9)  &    0.05(3)       &  0.6(4) & 0.03669 & 1.505   \\
                   &                     &                \hspace{13mm}   11/2$^{+}$   &    624.4          &   32(3)      &   0.11(6)    &    0.08(3)       &  1.5(6) & 0.05017 & 0.6245   \\
                   &                     &                \hspace{13mm}   13/2$^{+}$   &    653.6          &   100(3)     &   0.17(3)    &    0.22(9)       &  8(3)   & 0.1282  & 0.6027   \\
                   &                     &                \hspace{13mm}    7/2$^{+}$   &    859.5          &   22(3)      &              &       -           &  17(7) & -       & 2.817   \\
                   &                     &                \hspace{13mm}    9/2$^{+}$   &   1603.5          &   23(3)      &   (M1/E2)    &    $\ <$0.1      &  $\ <$1 & 0.00566 & 0.05378    \\ \\

   1665.6$^{IS}$  & 350$^{+100}_{-65}$  &  5/2$^{+}$     \hspace{5mm}     5/2$^{+}$   &    856.9          &  $\ <$1     &    (M1/E2)    &    $\ <$0.01     &         & 0.1079  & 0.3015   \\
                   &                     &                \hspace{13mm}     7/2$^{+}$  &    921.6          &  100(2)     &     1.4(2)    &    0.07(2)       & 90(35)  & 0.0740  & 2.891   \\
                   &                     &                \hspace{13mm}    9/2$^{+}$   &   1665.7$^*$      &   2(2)      &       E2      &        -          & $\ <$1 &  -      & 3.771  \\ \\

   1679.8$^{IS}$  & 315$^{+85}_{-60}$   &   7/2$^{+}$    \hspace{5mm}      5/2$^{+}$  &   364.1           &   60(3)     &  $-$0.17(9)   &    1.0(2)        & 130(50) & 0.00460 & 1.204     \\
                   &                     &                \hspace{13mm}     9/2$^{+}$  &   382.4           &   16(3)     &    (M1/E2)    &    $\ <$0.25     &         & 0.00001 & 1.150     \\
                   &                     &                \hspace{13mm}      5/2$^{+}$ &   870.1$^*$       &   7(3)      &    (M1/E2)    &    $\ <$0.01     &         & 0.00650 & 1.995     \\
                   &                     &                \hspace{13mm}      7/2$^{+}$ &   935.7           &  100(3)     &  0.09(9)      &    0.10(2)       &  0.5(1) & 0.00023 & 2.452     \\
                   &                     &                \hspace{13mm}      9/2$^{+}$ &  1679.7           &   36(3)     &   (M1/E2)     &   $\ <$0.01      &         & 0.00244 & 0.4702     \\ \\

   1683.3          & 150$^{+25}_{-20}$   &  9/2$^{+}$     \hspace{5mm}    9/2$^{+}$    &   600.7$^*$       &   17(4)     &   (M1/E2)     &   $\ <$0.15     &          & 0.0117  & 0.01288      \\
                   &                     &                \hspace{13mm}  11/2$^{+}$    &   704.2           &   42(4)     &  0.21(4)      &    0.20(4)      &   10(2)) & 0.05473 & 0.04624      \\
                   &                     &                \hspace{13mm}   7/2$^{+}$    &   939.3           &  100(4)     & $-$0.20(4)    &    0.21(3)      &   5.4(8) & 0.0962  & 0.6608     \\
                   &                     &                \hspace{13mm}   9/2$^{+}$    &  1683.2           &   54(4)     & $-$0.34(25)   &    0.02(1)      &   1(1)   & 0.00496 & 0.2582     \\ \\

   1686.3          & 250$^{+60}_{-40}$   &  13/2$^{+}$    \hspace{4mm}   11/2$^{+}$    &   707.4$^*$       &   88(4)     & $-$0.09(3)    &   0.20(5)   &  1.9(4)      & 0.01221 & 1.480    \\
                   &                     &                \hspace{14mm}  13/2$^{+}$    &   736.5           &   90(4)     & $-$0.27(13)   &   0.17(4)   &  13(3)       & 0.04217 & 0.06688     \\
                   &                     &                \hspace{14mm}   9/2$^{+}$    &  1686.3           &  100(4)     &      E2       &   -         &   3(1)       & -      & 0.1624     \\ \\

  1703.5$^{IS}$   & 220$^{+280}_{-90}$  &   3/2$^{+}$     \hspace{5mm}    5/2$^{+}$     &  387.9$^*$        &  100(4)     & $-$0.02(6)    &    2.4(2.0)   &  4(3)    & 0.02740 & 4.030    \\
                   &                     &                 \hspace{13mm}   5/2$^{+}$     &  894.8$^*$        &   87(4)     & $-$0.3(1)     &    0.15(12)   &  10(6)   & 0.5956  & 5.015    \\ \\

\hline \hline \\

\end{tabular}
\end{center}
\end{table*}



\begin{table*}[!]
\begin{center}
\renewcommand{\baselinestretch}{0.5}\small\small
\begin{tabular}{lccccccccccc}

\hline \hline \\
 E$_{level}$        &    $\tau$              &    $J^{\pi}_i$  $\rightarrow$  $J^{\pi}_f $        &    $E_{\gamma}$     & $I_{\gamma}$ &  $\delta$  &     $B(M1)$      &   $B(E2)$  & \hspace{3mm} $B(M1)$ & $B(E2)$ & IS & IV  \\
 (keV)              &     (fs)               &                                                    &   $(keV)$           &              &            &    ($\mu_N^2)$   &   (W.u.)   & \multicolumn{2}{c}{\hspace{3mm}[Shell Model]} &&  \\  \hline \\

   1773.1$^{IS}$   & 125$^{+20}_{-15}$      &   (1/2$^{+}$)     \hspace{5mm}        (5/2$^{+}$)  &   646.0$^*$       & 86(4)   &      E2          &                &           & -  &12.36  &  \\
                    &                        &                   \hspace{15mm}        5/2$^{+}$   &   964.0           & 100(4)  &      E2          &                &           & -  &5.126  &  \\ \\

   1812.2           & 150$^{+50}_{-35}$      &                   \hspace{15mm}       (17/2$^{+}$)  &   477.3           & 100     &                  &                &          &  \\ \\

   1910.6$^{IV}$   & 200$^{+40}_{-30}$      &   11/2$^{+}$   \hspace{5mm}        9/2$^{+}$       &   613.4$^*$       &  10(6)  &  $-$0.20(12       &    0.08(3)      &   5(2)  &0.00011  &  0.1171 &  \\
                    &                        &              \hspace{15mm}        9/2$^{+}$        &   828.1$^*$       &   7(6)  &  $-$0.61(17)       &    0.02(1)      &   6(3) & 0.04079 & 3.748 &    \\
                    &                        &              \hspace{15mm}        9/2$^{+}$        &  1910.6           & 100(6)  &   (M1/E2)        &   $\ <$0.03     &          & 0.00032 & 2.444 &      \\ \\

   1916.1$^{IV}$    &     90(10)             &    7/2$^{+}$   \hspace{6mm}       5/2$^{+}$        &   600.4$^*$       &  36(4)  &   0.06(4)        &   0.66(10)    &   4(1)     &  & &      \\
                    &                        &                 \hspace{14mm}      9/2$^{+}$       &   833.4           & 100(4)  &  $-$0.01(2)      &   0.69(10)    &  0.1(1)    &  & &      \\
                    &                        &                 \hspace{14mm}      5/2$^{+}$       &  1107.2$^*$       &   4(4)  &    (M1/E2)       &   $\ <$0.01   &            &  & &       \\
                    &                        &                 \hspace{14mm}      7/2$^{+}$       &  1172.1$^*$       &  12(4)  &    (M1/E2)       &   $\ <$0.03   &            &   & &          \\
                    &                        &                 \hspace{14mm}      9/2$^{+}$       &  1915.5$^*$       &   5(4)  &    (M1/E2)       &  $\ <$0.01   &             &  & &            \\ \\

   1949.6$^{IS}$   & 770$^{+1650}_{-320}$   &   7/2$^{+}$   \hspace{5mm}      (5/2$^{+}$)        &   270.1$^*$    & 100(5)  &      M1/E2        &              &               &   &&         \\
                    &                        &              \hspace{15mm}      9/2$^{+}$          &   866.8$^*$    &  9(5)   &      M1/E2        &              &               &   &&         \\
                    &                        &              \hspace{15mm}     11/2$^{+}$          &   971.1$^*$    &         &                   &              &               &   &&         \\
                    &                        &              \hspace{15mm}      5/2$^{+}$          &  1140.8        & 100(5)  &    0.21(5)        &              &               &   &&         \\
                    &                        &              \hspace{15mm}      7/2$^{+}$          &  1205.9        & 92(5)   &      M1/E2        &              &               &   &&         \\ \\

   1949.7$^*$       &  925$^{+3700}_{-425}$  &  (7/2$^{+}$)  \hspace{5mm}      (9/2$^{+}$)        &   266.4$^*$    &  26(4)   &   (M1/E2)       &              &                &  &&    \\
                    &                        &             \hspace{15mm}     11/2$^{+}$           &   346.4$^*$    &  41(4)   &     E2          &              &                &  &&    \\
                    &                        &             \hspace{15mm}      9/2$^{+}$           &  1949.8        &  82(4)   &   (M1/E2)       &              &                &  &&    \\ \\

   1968.3           &                        & (11/2$^{+}$,13/2$^{+}$)  \hspace{-1mm}    11/2$^{+}$ &   365.0$^*$    &  45(3)   &   (M1/E2)       &       &          &   &&    \\
                    &                        &                       \hspace{21mm}    15/2$^{+}$    &   477.3        & 100(3)   &   (M1/E2)       &       &          &  &&     \\    \\

   1968.8$^{IS}$    & 160$^{+35}_{-30}$      &    11/2$^{+}$         \hspace{3mm}    13/2$^{+}$     &   282.5        & 27(5)   &     (M1/E2)      &  $\ <$1.5     &          & &&      \\
                    &                        &                       \hspace{13mm}    9/2$^{+}$     &   285.6        & 34(5)   &     (M1/E2)      &  $\ <$1.8     &          & &&        \\
                    &                        &                       \hspace{13mm}   11/2$^{+}$     &   990.0        & 64(5)   &  $-$0.83(16)     &  0.05(2)      &  20(7)   &  &&        \\
                    &                        &                       \hspace{13mm}   13/2$^{+}$     &  1019.0        & 38(5)   &  $-$0.28(7)      &  0.04(1)      &   2(1)   &  &&         \\
                    &                        &                       \hspace{13mm}    7/2$^{+}$     &  1225.0$^*$    & 12(5)   &     E2           &               &   3(2)   &  &&          \\
                    &                        &                       \hspace{13mm}    9/2$^{+}$     &  1968.9        & 100(5)  &   (M1/E2)        &  $\ <$0.2     &          & &&           \\ \\

  2002.5$^{IS}$    &     $\ >$ 800          &  11/2$^{+}$    \hspace{5mm}    11/2$^{+}$    &    399.1$^*$    & 20(2)  &     (M1/E2)       &                &          &  &&      \\
                    &                        &              \hspace{15mm}    7/2$^{+}$      &    502.4$^*$    & 12(2)  &       E2          &              &          &     &&    \\
                    &                        &              \hspace{15mm}   15/2$^{+}$      &    511.5$^*$    &        &       E2          &              &          &    &&        \\
                    &                        &              \hspace{15mm}   11/2$^{+}$      &   1023.7$^*$    & 10(2)  &     M1/E2         &              &          &    &&      \\
                    &                        &              \hspace{15mm}   13/2$^{+}$      &   1052.8        & 100(2) &    $-$0.63(7)     &              &          &   &&    \\ \\

 2122.6$^{*,}$$^{IS}$  & 115$^{+30}_{-20}$ &    9/2$^{+}$    \hspace{5mm}     7/2$^{+}$       &  639.0$^*$     & 36(3)  &     (M1/E2)         &  $\ <$0.2     &          &0.09535  & 0.0023  &  \\
                        &                    &                 \hspace{15mm}   11/2$^{+}$       &  1143.7$^*$    & 71(3)  & 3.8$^{+1.9}_{-1.0}$ &   0.01(1)     &  29(6)   & 0.03691 &  2.275  &  \\
                        &                    &                 \hspace{15mm}    7/2$^{+}$       &  1378.9$^*$    & 29(3)  &  $-$1.9(8)          &   0.02(1)     &  0.2(2)  &0.00299  & 0.1651  &  \\ \\
                        &                    &                 \hspace{15mm}    9/2$^{+}$       &  2122.6$^*$    & 100(3) &     (M1/E2)         &   $\ <$0.02   &          & 0.00078 &1.508    &  \\ \\

   2162.6$^{IS}$   &  410$^{+300}_{-125}$   &   (13/2$^{+}$)  \hspace{5mm}    15/2$^{+}$       &   671.7$^*$    & 25(4)  &   (M1/E2)           &               &          &   &&       \\
                    &                        &                 \hspace{15mm}   11/2$^{+}$       &  1183.7        & 100(4) &   (M1/E2)           &               &          &   &&       \\
                    &                        &                 \hspace{15mm}   13/2$^{+}$       &  1212.8$^*$    & 61(4)  &   (M1/E2            &               &          &  &&        \\ \\

   2170.4$^{IS}$   &  350$^{+165}_{-90}$    &    9/2$^{+}$   \hspace{5mm}     9/2$^{+}$        &  1087.6$^*$    & 10(3)  &   (M1/E2)           &               &          &  &&    \\
                    &                        &                \hspace{15mm}   11/2$^{+}$        &  1192.5        & 100(3) &   (M1/E2)           &               &          &   &&   \\
                    &                        &                \hspace{15mm}   13/2$^{+}$        &  1221.6        & 60(3)  &     E2              &               &          &   &&    \\
                    &                        &                \hspace{15mm}    5/2$^{+}$        &  1361.1$^*$    & 31(3)  &     E2              &               &          &   &&   \\
                    &                        &                \hspace{15mm}    7/2$^{+}$        &  1426.1$^*$    & 27(3)  &   (M1/E2)           &               &          &    &&  \\ \\

   2184.0$^*$       &  110$^{+45}_{-30}$     &   (13/2$^{+}$) \hspace{5mm}   (17/2$^{+}$)       &   849.1$^*$    & 100(2) &     E2              &               &          &  &&     \\
                    &                        &                \hspace{15mm}  (9/2$^{+}$)        &   480.5$^*$     & $\ <$2 &    E2              &               &          &  &&     \\ \\

\hline \hline
\end{tabular}
\end{center}
\end{table*}


\subsection{Particle-core ($PC$) weak coupling model}

The states of an odd-A nucleus can be described in terms
of a valence nucleon coupled to the excited
states of the neighboring even-even core
\cite{unifiedmodel,bm2,choudhury}. In the weak-coupling limit, the
coupling Hamiltonian can be treated as a perturbation, whereas the
intermediate coupling also allows for mixing between several
single-particle states coupled to core excitations. The angular
momentum of the core, J$_{core}$ and single-particle (or particle-hole) states,
J$_{sp}$, couple to form a multiplet of states with a total angular
momentum
\cite{core-excitations,centerofgravity,particle-vibration,87y,93nb_old}
given by,
\begin{equation}
\mid J_{sp}-J_{core}\mid \hspace{1mm} \leq J_{core}\otimes J_{sp}
\hspace{1mm} \leq \hspace{1mm} \mid J_{sp}+J_{core}\mid.
\end{equation}
\noindent In fact, low-lying excited states in $^{93}$Nb can be  regarded as
resulting from the coupling of a $1g_{9/2}$ proton  to a
$^{92}_{40}$Zr core and a $2p_{1/2}^{-1}$ proton-hole to a
$^{94}_{42}$Mo core \cite{93nb_old}. These couplings
result in two independent and unmixed one-phonon structures of
opposite parity: a) a quintet of positive-parity states  built on
the J$^\pi$=9/2$^+$ ground state, resulting from the $\pi
1g_{9/2}$$\otimes$(2${^+_1}$,$^{92}$Zr) $PC$ coupling;
and b) a pair of negative-parity states built on the isomeric (with a half-life of
16 years) J$^\pi$=1/2$^-_1$ state at 31 keV, which corresponds to the $\pi
2p_{1/2}^{-1}$$\otimes$(2${^+_1}$,$^{94}$Mo) configuration.

\subsection{$\pi 1g_{9/2}$$\otimes$(2${^+_1}$,$^{92}$Zr) configuration}

The $\pi 1g_{9/2}$$\otimes$(2${^+_1}$,$^{92}$Zr) configuration assignment is supported by the center-of-gravity
theorem \cite{centerofgravity,93nb_old}, which implies, through the j-j coupling shell model, the existence
of geometrical relations among the spectra of neighboring nuclei. The center-of-gravity energy of the
one-phonon system, $\Delta E_{CG}$, is then given by the  relation,
\begin{equation}
(2j_{p}+1)\Delta E_{CG}=(2J_{core}+1)^{-1} \sum_{J_3}(2J_3+1)E_{J_3}
\label{cg}
\end{equation}
\noindent where  $j_p$ is the angular momentum of the coupled
particle (in our case $g_{9/2}$), $J_{core}$ is the angular momentum
of the core state (2$^+$), and $J_3$, $E_{J_3}$ are the spins and
excitation energies, respectively, of the single one-phonon states.
Considering the first 5 excited positive-parity states in $^{93}$Nb
as  one-phonon excitations and Z=40 as a semi-closed
shell for protons, the predicted center-of-gravity excitation energy
is 934 keV. This energy is in striking agreement with the 934.5 keV
measured for the first excited 2$^+$ state in $^{92}$Zr
\cite{fransen92zr,bunker}. In addition, from Coulomb excitation
studies, the sum of $B(E2)$$\uparrow$ values for the quintet of
one-phonon states proposed in $^{93}$Nb, 765(11) e$^2$ fm$^4$
(weighted average from Refs.~\cite{kregar, stelson, yoshizawa, kakavand}),
matches well the excitation of the 2$^+$ core state, 795(56) e$^2$ fm$^4$
(weighted average from Refs.~\cite{alder, yoshizawa}), in $^{92}$Zr.
Again, these data support the weak-coupling nature of these states
\cite{kregar, alder}.

However, the above arguments are not consistent with  the results from 
Kent {\it et al.} in the positive-parity structure, where inelastic proton scattering studies through isobaric analog
resonances in $^{94}$Mo did not support the weak
coupling in $^{93}$Nb~\cite{kent}. The proposed
9/2$^+_2$ member of the quintet at 1082.6 keV decays preferentially to
the 7/2$^+_1$ state at 744.0 keV rather than by a  1082.6 keV
transition to the ground-state; which breaks the selection
rule between vibrational states ($\Delta n_{\lambda}=\pm1$).
Two-state mixing calculations between the 1082.6-keV and ground 
states were done in an attempt to explain this anomalous decay
\cite{kregar}, and the results are in agreement with decay strengths
in neighboring $^{92}$Mo and $^{94}$Mo nuclei.



\subsection{$\pi 2p_{1/2}^{-1}$$\otimes$(2${^+_1}$,$^{94}$Mo) configuration}

Low-lying negative-parity states in $^{93}$Nb  can be regarded as resulting
from the $\pi2p_{1/2}^{-1}$$\otimes$(2${^+_1}$,$^{94}$Mo) coupling. A
doublet of negative-parity states (J$^\pi$=3/2$^-$ at 687.4 keV and J$^\pi$=5/2$^-$ at 810.7 keV)  
built on the J$^\pi$=1/2$^-_1$  state confirms this
assignment. Recently, we have studied excited states built on this
doublet and identified isovector excitations that correspond to the 2$_{1,IV}^+$
states found in neighboring even-even nuclei \cite{orce}.
Identifications are based on $M1$ and $E2$ strengths, energy
systematics, and spin-parity assignments and from the comparison
with $SM$ calculations with the low-momentum nucleon-nucleon
interaction, $V_{low-k}$ \cite{bogner03}. Similar investigations
will be provided in this work for the $IS$ negative-parity
excitations, where seven states are expected from the $PC$ model, i.e., 
in addition to the 1/2$^-_1$ state for the single-particle state,
3/2$^-$ and 5/2$^-$ states for the one-phonon excitations, and
3/2$^-$, 5/2$^-$, 7/2$^-$, and 9/2$^-$ states for the two-phonon excitations.

\subsection{Positive-parity states}

Five new levels and 40 additional $\gamma$-ray transitions have been identified in
this work. Table~\ref{tab:pos} lists the positive-parity states below 2.2 MeV, together with the $\gamma$ rays
depopulating them. The decay properties evince a complex structure of
levels decaying to either the ground state or lowest $IS$ excitations. 
Despite possible admixtures of $IS$ and $IV$ wavefunctions in the odd-mass case,
corresponding $MS$ decay signatures might be expected analogous to those observed in the even-A $N=52$ isotones
\cite{9,fransen}. From strong $M1$ transitions  to
the $IS$ states and weakly collective $E2$ transitions to the ground
state, we propose three candidates for $IV$ excitations, the 9/2$^+_2$,  5/2$^+_2$ and 7/2$^+_3$ levels at 
1297.4, 1315.7 and 1483.6 keV, respectively. We propose them to be members of the quintet
of $IV$ excitations (5/2$^+_{IV}$, 7/2$^+_{IV}$, 9/2$^+_{IV}$, 11/2$^+_{IV}$ and 13/2$^+_{IV}$)
arising from the $\pi 1g_{9/2}$$\otimes$(2$_{1, {\rm IV}}^{+}$,$^{92}$Zr) coupling.
Furthermore, from enhanced $B(E2)$ values to the  9/2$^+_{IV}$ state, we tentatively propose the 11/2$^+$ level at 1910.6 keV
as a fragment of the scissors mode. Additional details are provided below. 
Further assignments in Table \ref{tab:pos} are tentative due to the complex structure and possibility 
of intermediate coupling.

\subsubsection{1297.4 keV 9/2$^+_{IV}$ state}

The 1297.4 keV level has  been firmly assigned as  J$^{\pi}$=9/2$^+$
from the analysis of the angular correlation of the decay branches
(see Table \ref{tab:pos} and the bottom panel of Fig.
\ref{fig:1}). A  mean lifetime of 380$^{+110}_{-75}$ fs has been determined for this
level from the Doppler-shift attenuation data. This
value is in agreement with, but more accurate than the 300$^{+300}_{-100}$ fs lifetime
measured in previous work~\cite{russian}. Hence, the 553.1 and 318.3 keV transitions have
large $B(M1)$ values of 0.33(5)  and $\ <$0.9 $\mu{^2_N}$,
respectively. This level also decays by a small E2 strength to the ground state, 
$B(E2; 9/2^+_{IV} \rightarrow 9/2^+_1)=1.3(3)$  W.u. Both features
are typical signatures for $MS$ states.

\subsubsection{1315.7 keV 5/2$^+_{IV}$ state}
A  J$^{\pi}$=5/2$^+$ assignement has been given to this level from the 
angular correlation of the decay branches (see Table \ref{tab:pos}) together with 
a mean lifetime of 530$^{+450}_{-170}$ fs from the Doppler-shift attenuation data. 
The 571.5-keV transition to the 7/2$^+_1$ has a large $B(M1)$ value of 0.45(25) $\mu{^2_N}$; nonetheless, the 506.7-keV 
transition to the 5/2$^+_1$ state presents a large $B(E2)$ value. Hence, the excitation has not a pure $IV$ character and 
intermediate coupling may mix $IS$ and $IV$ states.

\subsubsection{1483.6 keV 7/2$^+_{IV}$ state}

The  1483.6 keV level  was previously assigned as
J$^{\pi}$$=$(7/2$^+$, 9/2$^+$). The angular correlation analysis of
the  branches depopulating this state (see Table
\ref{tab:pos}) has firmly  assigned it as  J$^{\pi}$=7/2$^+$.
A  short mean lifetime of 65(5) fs has been determined for the first
time. The 674.8- and 400.8-keV transitions have large $B(M1)$ values
of 0.58(6) and $\ <$0.7 $\mu{^2_N}$, respectively. The level decays
to the ground state through a weakly collective E2 transition,
$B(E2)=0.9(6)$ W.u. Again, we find the charateristic features of a
$MS$ state.

\subsubsection{1910.6 keV $11/2^+$ state}

The 1910.6 keV level has been assigned as J$^{\pi}$$=$11/2$^+$, in
disagreement with previous work, where  J$^{\pi}$$=$7/2$^{(+)}$ was
proposed \cite{russian2}. A lifetime of 200$^{+40}_{-30}$ fs has been
determined  for this level, which decays with a weakly collective
$B(E2)$ value of $5(2)$ W.u. to the proposed 1297.4 keV $IV$ excitation.
From this  B(E2) value and the spin assignment of the
state, we tentatively propose the 1910.6 keV level as a fragment of
a second-order $IV$ excitation identified in the even-even neighbors.
Arguments against this assignment are that neither the 1910.6 keV
transition to the ground state nor the 828.1-keV $\gamma$ ray to the
9/2$^+_2$  state  have the enhanced $B(M1)$ character
expected for an $IV$ excitation, and the scissors mode is generally
identified at about 3 MeV. Nevertheless, the $IV$
assignment is plausible since large fragmentation of the scissors
mode strength has been observed in the 2 to 4 MeV  energy range  in
systematic studies of odd-A rare earth nuclei
\cite{enders,nord,nord2}.

\subsubsection{1968.8 keV $11/2^+_{IS}$ state}

Although the spin was assigned as J$^{\pi}$$=$11/2$^+$ and we have a
newly determined lifetime of 160$^{+35}_{-30}$ fs for this state, only upper
values for the $M1$ strengths of some of the transitions could be
determined. The large $B(E2)=20(7)$ W.u. observed for the 990.0-keV
transition to the 11/2$^+_1$ $IS$ excitation is noteworthy.

\subsubsection{2122.6 keV $9/2^+_{IS}$  state}

With an assigned J$^{\pi}$$=$9/2$^+$ and a  lifetime of 115$^{+30}_{-20}$  fs,
this state presents small $B(M1)$ values ($\ <$0.2 $\mu_N^2$) and a
relatively large $B(E2)=29(6)$ W.u. to the
11/2$^+_1$ isoscalar excitation.

\begin{table*}[]
\begin{center}
\caption{Properties of low-lying negative-parity states in $^{93}$Nb
up to 2.1 MeV. Level and $\gamma$-ray energies, branching and mixing
ratios, initial and final spin of the states, lifetimes, and reduced transition probabilities are listed. An asterisk labels newly
identified levels and $\gamma$ ray transitions. Shell model $B(M1)$
and $B(E2)$ predictions for relevant transitions are shown on the
right with the isoscalar (IS) and isovector (IV) components of the
$E2$ operator.  Uncertainties in the energies
are 0.2 keV.} \label{tab:strengthsnega}
\begin{tabular}{lcccccccccccc}
\hline \hline
$E_L$       & \hspace{3mm} $\tau$  & &  $J^\pi_i$ $\rightarrow$ $J^\pi_f$ & \hspace{3mm} $E_\gamma$  & $I_\gamma$ & \hspace{3mm} $\delta$  & \hspace{3mm} B($M1$)   & \hspace{3mm} B($E2$)        & \hspace{3mm} $B(M1)$ & $B(E2)$ & IS & IV \\

   (keV)    & \hspace{3mm} (fs)    & &                                    & \hspace{3mm} (keV)       &            & \hspace{3mm}           & \hspace{3mm} $(\mu_N^2)$   & \hspace{3mm}  (W.u.)    &  \multicolumn{2}{c}{\hspace{3mm}[Shell Model]} && \\ \hline

30.9       &                         &  &  $\frac{1}{2}^-$
\hspace{3mm}    $\frac{9}{2}^+$  &                  &          &
$M4$        &                         &      & & & & \\ [0.7ex]

687.4        &  400$^{+70}_{-20}$$^q$  &  &  $\frac{3}{2}^{-}$
\hspace{3mm} $\frac{1}{2}^-$  &    655.9         &   100    &
(M1/E2)   &                         &      & & & &  \\ [0.7ex]

810.7       &                         &  &  $\frac{5}{2}^{-}$
\hspace{3mm}    $\frac{1}{2}^-$  &    799.6         &   100    & E2
&                         &      & & & &  \\ [0.7ex]
                &                         &  &                     \hspace{8.7mm}  $\frac{3}{2}^-$  &    123.3$^*$     &  $\ <1$  &    (M1/E2)   &                         &      & & & &  \\[0.7ex]

1284.8     &  $250^{+80}_{-50}$    &  &  $\frac{5}{2}^-$
\hspace{3mm}    $\frac{1}{2}^-$  &  1253.5          &  100(4)  &
$E2$       &                         & $32^{+10}_{-9}$    &- &-&-&-
\\ [0.7ex]

               &              &  &                           \hspace{8.7mm}  $\frac{3}{2}^-$  &   597.3$^*$      &   25(4)  &  $0.14(4)$   & $0.20^{+0.10}_{-0.08}$  & $6^{+3}_{-2}$      & -&-&-&- \\[0.7ex]

               &              &  &                           \hspace{8.7mm}  $\frac{5}{2}^-$  &   473.9$^*$      &    5(4)  &   $(M1)$     &    $\ <$0.08            &                    &  -&-&-&- \\[0.7ex]

1370.1$^{IS}$      &   $\ > 790$  &  &  $\frac{5}{2}^-$ \hspace{3mm}
$\frac{3}{2}^-$  &  683.2$^*$   &   30(4)  &  $-0.34(5)$     &  $\
<$0.05           &  $\ <$7         & -&-&-&-  \\ [0.7ex]

               &              &  &                  \hspace{8.7mm}  $\frac{5}{2}^-$  &  559.4       &  100(4)  &  $-0.32(7)$     &  $\ <$0.29            &  $\ <$54&  -&-&-&- \\ [0.7ex]

1395.8$^{IS}$     &   $\ > 790$  &  &  $\frac{7}{2}^-$ \hspace{3mm}    $\frac{3}{2}^-$  &   708.6      &   9(4)  & $E2$             &                  &    $\ <$18            &- & 8.50 & 28.61 & -13.85  \\[0.7ex]

               &              &  &                  \hspace{8.7mm}  $\frac{5}{2}^-$  &   585.1      & 100(4)  & $-0.10(2)$       &    $\ <$0.31     &   $\ <$5.2                & 0.00003 & 0.879 & 9.17  & -4.39  \\[0.7ex]

1500.0$^{IS}$     &  1170(300)$^{\dagger}$  &  &  $\frac{9}{2}^-$  \hspace{3mm}  $\frac{5}{2}^-$    &   689.5(1)     &  18(3)      & $E2$  &             &   $27^{+15}_{-9}$    &  -  & 10.63 & 35.39  & -15.21  \\[0.7ex]

1572.1$^{IS}$     &   $280^{+210}_{-100}$    &  &  $\frac{3}{2}^-$
\hspace{3mm}    $\frac{3}{2}^-$  &  885.1$^*$  &   37(5)   &
$-1.60(14)$      &   $0.02(1)$                 &  $36^{+27}_{-19}$
&0.002 & 14.46 & 25.71  & -8.85  \\ [0.7ex]

               &                         &  &                  \hspace{8.7mm}  $\frac{5}{2}^-$  &  761.4$^*$  &  100(5)   &    $-0.28(3)$       &   $0.27^{+0.16}_{-0.12}$    &  $21^{+12}_{-10}$ & 0.00001 & 6.0 & 16.66 & -6.44 \\[0.7ex]

               &                         &  &                  \hspace{8.7mm}  $\frac{5}{2}^-$  &  287.4$^*$  &   20(5)   &      $(M1)$         &   $\ <$1.09                          &                     \\[0.7ex]

1588.4$^{IS*}$  &   $\ > 1260$            &  &  $\frac{5}{2}^-$
\hspace{3mm}    $\frac{3}{2}^-$  &  901.2$^*$  & 100(8)   &
$-0.53(6)$                  &  $\ <$0.04     &     $\ <$8  & 0.0012
& 4.36 & 17.18& -5.26  \\ [0.7ex]

               &                         &  &                  \hspace{8.7mm}  $\frac{5}{2}^-$  &  777.8$^*$  &  18(8)   &    $-4.03^{+1.32}_{-3.45}$    &  $\ <$0.001     &     $\ <$13 & 0.0069 &16.94 &34.19 & -12.37 \\[0.7ex]

1779.7$^{IV}$      &   $105^{+43}_{-28}$  &  & \hspace{-1.2mm}
$\frac{5}{2}^{(-)}$ \hspace{1mm}    $\frac{3}{2}^-$  &  1092.4$^*$ &
8(5)   & $0.05(9)$     &  $0.03^{+0.04}_{-0.02}$     &
$0.04^{+0.04}_{-0.03}$ &0.037 & 0.306 &-0.74 & 2.35  \\ [0.7ex]

               &                      &  &                  \hspace{8.7mm}  $\frac{5}{2}^-$  &   969.0      & 100(5) & $0.04(6)$     &  $0.55^{+0.24}_{-0.18}$     &  $0.5(2)$ & 0.616 & 0.0001 &-0.72 & 4.92 \\[0.7ex]

               &                      &  &                  \hspace{8.7mm}  $\frac{1}{2}^-_{gs}$  &   & & & & & - & 4.28 & 12.63 & 19.87 \\[0.7ex]

1840.6$^{IV*}$  &   $103^{+35}_{-24}$   &  & \hspace{-4mm}  $\frac{3}{2}^{-}, \frac{5}{2}^-$ \hspace{0mm}    $\frac{3}{2}^-$  &  1153.4$^*$  & 100(4)  & $0.14(4),0.26(6)$     &  $0.29(8),0.28(9)$     &  $2.5(7),8(2)$ & 0.462 & 0.306 & -4.36  & 4.90 \\

               &                       &  &                                                 \hspace{8.7mm}  $\frac{5}{2}^-$  &  1029.6$^*$  & 20(4)   & $0.32(6),0.17(8)$     &  $0.08(3),0.08(3)$     &  $4(1),1(1)$ & 0.046 & 0.078 & -1.91 & 2.63 \\[0.7ex]

 & & & \hspace{8.7mm}  $\frac{1}{2}^-_{gs}$ & &&&& & - & 5.99 & 13.03 & 14.55 \\[0.7ex]

1948.1     &  $230^{+133}_{-69}$  &  & \hspace{-1.2mm}  $\frac{7}{2}^{(-)}$ \hspace{1mm}    $\frac{5}{2}^-$  &  1137.4      & 100    & $0.05(4)$  &  $0.17(7)$     &  $0.2(1)$\\[0.7ex]

1997.6$^*$  &   $92^{+22}_{-17}$   &  &  $\frac{5}{2}^-$ \hspace{3mm}    $\frac{3}{2}^-$  &  1310.2$^*$  & 12(5)  & $-0.29(12)$   &  $0.03(2)$     &  $0.8^{+0.6}_{-0.4}$\\[0.7ex]

               &                      &  &                  \hspace{8.7mm}  $\frac{5}{2}^-$  &  1186.9$^*$  & 100(5) & $-0.31(11)$   &  $0.30^{+0.09}_{-0.07}$     &  $12^{+4}_{-3}$\\[0.7ex]

2012.6$^*$  &                      &  &  $\frac{3}{2}^-$  \hspace{3mm}    $\frac{3}{2}^-$  &  1325.8$^*$  & 100(5)    & $4.47^{+1.53}_{-0.94}$ &     &  \\

               &                      &  &                   \hspace{8.7mm}  $\frac{3}{2}^-$  &  440.4$^*$   &   6(5)    &                        &     &  \\[0.7ex]

2024.4$^{IS*}$  &  $78^{+41}_{-25}$    &  & $\frac{3}{2}^-$   \hspace{3mm}    $\frac{3}{2}^-$  &  1337.1$^*$  &  100(3)    & $-4.70^{+0.84}_{-1.28}$   & $0.01(1)$      &  $91^{+76}_{-53}$    \\[0.7ex]

               &                      &  &                   \hspace{8.7mm}  $\frac{3}{2}^-$  &  452.1$^*$   &   3(3)     &     $(M1)$                & $\ <0.23$      &  \\[0.7ex]

2099.6     &  $133^{+62}_{-36}$    &  & \hspace{-1.4mm}  $\frac{7}{2}$  \hspace{5.3mm}   $\frac{5}{2}^-$  &  1288.9$^*$ &  46(7)    & $-0.05(5)$    &  $0.05(1)$     & $0.2^{+0.3}_{-0.1}$      \\

              &                       &  &                       \hspace{8.7mm} $\frac{7}{2}^-$  &   703.8     & 100(7)    &               &                &                          \\[0.7ex]

2127.1$^*$  & $235^{+171}_{-109}$   &  &         \hspace{8.7mm}   $\frac{5}{2}^-$  &  1383.1$^*$   &  13(4)    &     &       &       \\
               &                       &  &         \hspace{8.7mm}   $\frac{5}{2}^-$  &  1316.61$^*$  &  10(4)    &     &       &       \\
               &                       &  &         \hspace{8.7mm}   $\frac{7}{2}^-$  &   731.3$^*$   &  18(4)    &     &       &       \\
               &                       &  &         \hspace{8.7mm}   $\frac{7}{2}^+$  &   626.9$^*$   & 100(4)    &     &       &       \\[0.7ex]

2153.8$^*$  & $115^{+28}_{-20}$     &  &         \hspace{8.7mm}   $\frac{1}{2}^-$  &  2122.8$^*$   &  100      &     &       &       \\[0.7ex]

 \hline \hline
\end{tabular}
\end{center}
\end{table*}

\begin{table*}[!]
\begin{center}
\vspace{-0.7cm}\hspace{-12.3cm}
\begin{tabular}{lccccccccccccc}
$^\dagger$ Data taken from Ref. \cite{stelson}.\\
$^q$ Lifetime taken from Ref. \cite{coral}.\\\\
\end{tabular}
\end{center}
\end{table*}


\subsection{Negative-parity states}

Although some relevant information concerning the negative-parity states in $^{93}$Nb has already been
published~\cite{orce}, we present additional data collected in our measurements. Table \ref{tab:strengthsnega} 
lists the results for levels up to 2.1 MeV.

\subsubsection{1284.8  5/2$^-$  state}

The 1284.8 keV level has been assigned as J$^{\pi}$$=$5/2$^-$. A
lifetime of 250$^{+80}_{-50}$ fs has been  measured for this level,
which decays through a large $B(E2)$ value of 32$^{+10}_{-9}$ W.u.
to the 1/2$^-$ single-particle state, indicating a strong
correlation between the wavefunctions of these states.

\subsubsection{1370.1  5/2$^-_{IS}$ state}

The 1370.1 keV level has been assigned as J$^{\pi}=5/2^-$. A
lower limit for the lifetime of $>$790 fs has been determined for
this level, giving upper limits for the B(E2) values to the 3/2$^-_1$
and 5/2$^-_1$ one-phonon states of  $<$7 and $<$54 W.u., respectively. No
decay to the ground state has been observed,  supporting its
$IS$ character.

\subsubsection{1395.8 7/2$^-_{IS}$  state}

The 1395.8 keV level has been assigned as J$^{\pi}=7/2^-$. A
lower limit for the lifetime of $>$790 fs has been determined for
this level, giving upper limits for the B(E2) values to the 3/2$_1^-$
and 5/2$_1^-$ states of  $<$18 and $<$5 W.u., respectively. No
decay to the ground state has been observed, supporting its
$IS$ character.  In addition, the strong isoscalar character
of the transitions to the one-phonon states, listed in the last two
columns of Table~\ref{tab:strengthsnega}, supports its $IS$
assignment.

\subsubsection{1500.0 keV 9/2$^-_{IS}$  state}

This state was previously assigned as J$^{\pi}=7/2^{(-)}$ \cite{russian2}.
Nevertheless, a J$^{\pi}=9/2^-$ assignment is a better solution from
the angular correlation fits. It also presents the longest lifetime,
a newly measured 1170(300) fs, with respect to the other levels
discussed in this section. An enhanced $B(E2)$ value of
27$^{+15}_{-9}$ W.u. for the  689.6 keV $E2$ transition to the
5/2$^-$ one-phonon state suggests this state as a member of the
negative-parity $IS$ coupling excitations. This large B(E2) value is
also predicted by our $SM$ calculations, together with the 
isoscalar character of the transition to the one-phonon
states.

\subsubsection{1572.1 3/2$^-_{IS}$ state}

A  lifetime of
280$^{+210}_{-100}$ fs has been determined for this level, giving
large  B(E2) values to the 3/2$^-$ and 5/2$^-$ one-phonon states of
36$^{+27}_{-19}$ and 21$^{+12}_{-10}$ W.u., respectively. These
large B(E2) values support the $IS$ character of the state.
Decay to the ground state has not been observed. The 
strong $E2$ transitions to the one-phonon
states confirm the $IS$ character.

\subsubsection{1588.1 5/2$^-_{IS}$  state}

A lower limit
for the lifetime of $>$1260 fs has been determined for this level,
giving upper limits for the B(E2) values to the 3/2$^-$ and 5/2$^-$
one-phonon states of  $<$8 and $<$13 W.u., respectively. Although the
experimental data are not conclusive, our $SM$ calculations
predict relatively strong transitions and isoscalar character for
the transitions to the one-phonon states, which  support the
$IS$ character.

\subsubsection{1779.7 keV 5/2$^-_{IV}$ state}

Although the identification of $MS$ states have already been discussed
in \cite{orce}, we include them  for completeness.
The previously proposed (5/2$^-$) level at 1779.7 keV yields a  new
1092-keV branch to the first 3/2$^-_1$ excited state that has been
revealed  from the excitation function and coincidence data. The
level has been assigned as  J$^{\pi}$=5/2$^{(-)}$ by the analysis of
the angular correlation data, and a mean lifetime of
105${^{+43}_{-28}}$ fs has been determined~\cite{orce}. The 969 and
1092 keV transitions depopulating this state to the
$2p_{1/2}^{-1}\otimes2^+$ symmetric one-phonon states provide
branching ratios of 100(5) and 8(5), respectively, and mixing
ratios, $\delta$, of $0.04(6)$ and $0.05(9)$, respectively. Hence,
the 969 keV transition to the 5/2$^-_1$ state has a large $B(M1)$
value of 0.55${^{+0.24}_{-0.18}}$ $\mu{^2_N}$ and a small $B(E2)$
value of 0.5(2) W.u., while the 1092 keV transition to the 3/2$^-_1$
state exhibits a much weaker $B(M1)$ strength of
0.03${^{+0.04}_{-0.02}}$ $\mu{^2_N}$ and a $B(E2)$ value of
0.04${^{+0.04}_{-0.03}}$  W.u. A large B(M1) value to the 5/2$^-$
one-phonon state was predicted, together with  strong isovector
character for such a transition.

\subsubsection{1840.6 keV 3/2$^-_{IV}$ state}

 The  1840.6 keV level has been placed
from our measurements. The angular correlation analysis of the
competing branches depopulating this state (see Table
\ref{tab:strengthsnega})
 leads equally to  either  J$^{\pi}$=3/2$^-$ or 5/2$^-$ assignments.
A  mean lifetime of 103$^{+35}_{-24}$ fs has been measured for this
state, yielding large $B(M1)$ values of 0.29(8) $\mu{^2_N}$
(J$^{\pi}$=3/2$^-$) or 0.28(9) $\mu{^2_N}$ (J$^{\pi}$=5/2$^-$) for
the 1153 keV transition. We proposed, however, that this state is
3/2$^-$, based on its proximity to the 1779.7 keV level and their 
rather different decay strengths. A large $B(M1)$ value, this time to
the 3/2$^-$ one-phonon state was predicted.

The $B(M1)$ values from the 1779.7-keV state to the 5/2$^-_1$ level
and from the proposed 1840.6 keV state to the 3/2$^-_1$ level are
greater than from any other negative-parity states feeding the
symmetric one-phonon states. These observations, together with the
appearance of these states in the expected energy range ($\sim$ 2
MeV), support their assignment as first-order isovector excitations.

The increasing level density and  stronger mixing above
1.9 MeV make the attempt of characterization of other excitations 
unrewarding.

\section{Discussion}

\subsection{Shell-Model Calculations}

For the present work, we solve the model space Schr\"odinger
equation $PH_{eff}P\Psi=EP\Psi$, where $H_{eff}=H_0+V_{eff}$ and
$V_{eff}$ is the $SM$ effective interaction. To derive
$V_{eff}$ we use the model space folded-diagram methods detailed in
Ref.~\cite{kuo90}, where we have the explicit expansion for the effective
interaction:
\begin{eqnarray}
\nonumber
V_{\rm eff} &=& \hat{Q} - \hat{Q'} \int \hat{Q} + \hat{Q'} \int \hat{Q} \int
\hat{Q} \\
&& - \hat{Q'} \int \hat{Q} \int \hat{Q} \int \hat{Q}+ \cdots.
\label{qbox}
\end{eqnarray}
In this series, $\hat{Q}$ represents the irreducible vertex
function, consisting of irreducible valence-linked diagrams, and the
integral sign denotes a generalized folding operation. In the
$\hat{Q}$-box we have included core polarization diagrammatic
contributions up to second order, which has been shown to be a good
approximation to a non-perturbative all-order summation in the absence
of $3N$ forces \cite{holt05}. The intermediate particle and
hole states are allowed two oscillator shells above and
below the model space (which is discussed below).  We sum the above
series using the Lee-Suzuki iteration method \cite{leesuz}.

Starting from any high-precision NN interaction {\vnn}, we can derive a
resolution-dependent low-momentum interaction {\vlk}, which preserves all
low-energy data below the chosen momentum cutoff \cite{bogner03}.  For this
work  we have used the
CD-Bonn interaction as {\vnn} with a momentum cutoff value of
$\Lambda=$2.0 {\fm}.
 There have now been a number of nuclear structure studies using this
low-momentum NN interaction in this nuclear region to describe a
variety of nuclei and their observables \cite{orce,holt07,burda07,werner08}.

Following the prescription of Ref.~\cite{3}, we choose to use
$\rm ^{88}Sr$ as the inert core for these calculations. Valence
neutrons can then occupy the following single-particle orbits above
the $N=50$ closed shell: $g_{7/2}$, $d_{5/2}$, $d_{3/2}$, $s_{1/2}$,
and $h_{11/2}$. For the valence protons, we take the $p_{1/2}$
and $g_{9/2}$ orbits consistent with a $Z=38$ proton
core. The single-particle energies, $\epsilon_j$, were obtained from
the experimental values in the $^{89}$Y and $^{89}$Sr nuclei; for
reference, they are listed in Table~\ref{tab:spe}.
These single-particle energies differ slightly
from those used in Ref.~\cite{orce}, which were tuned to best reproduce 
the experimental
spectra in $^{90}$Zr and $^{90}$Sr.  To
keep the calculations as free from adjustable parameters as possible,
we now use the experimental values, but our current results on the
negative-parity states in {\nb}
are essentially unchanged from those reported earlier~\cite{orce}.

\begin{table}
\centering
\caption{Single-particle energies for the orbits used in  $SM$
calculations.}
\vskip.1in
\label{tab:spe}
\begin{tabular}{lcccccc}
\hline \hline
\\
Proton orbits: & $p_{1/2}$ & $g_{9/2}$ & & & &  \\ [.05in]
Energy (MeV) & -0.91 & 0.0 & & & &  \\ [.1in]\hline
\\
Neutron orbits:& $g_{7/2}$ & $d_{5/2}$ & $d_{3/2}$
& $s_{1/2}$  & $h_{11/2}$ &  \\ [.05in]
Energy (MeV) & 1.47 & 0.0 & 2.01 & 1.03 & 3.00 &  \\[.1in] \hline \hline
\end{tabular}
\end{table}

\begin{figure}[t]
\begin{center}
\includegraphics[width=8cm,height=8cm,angle=-90]{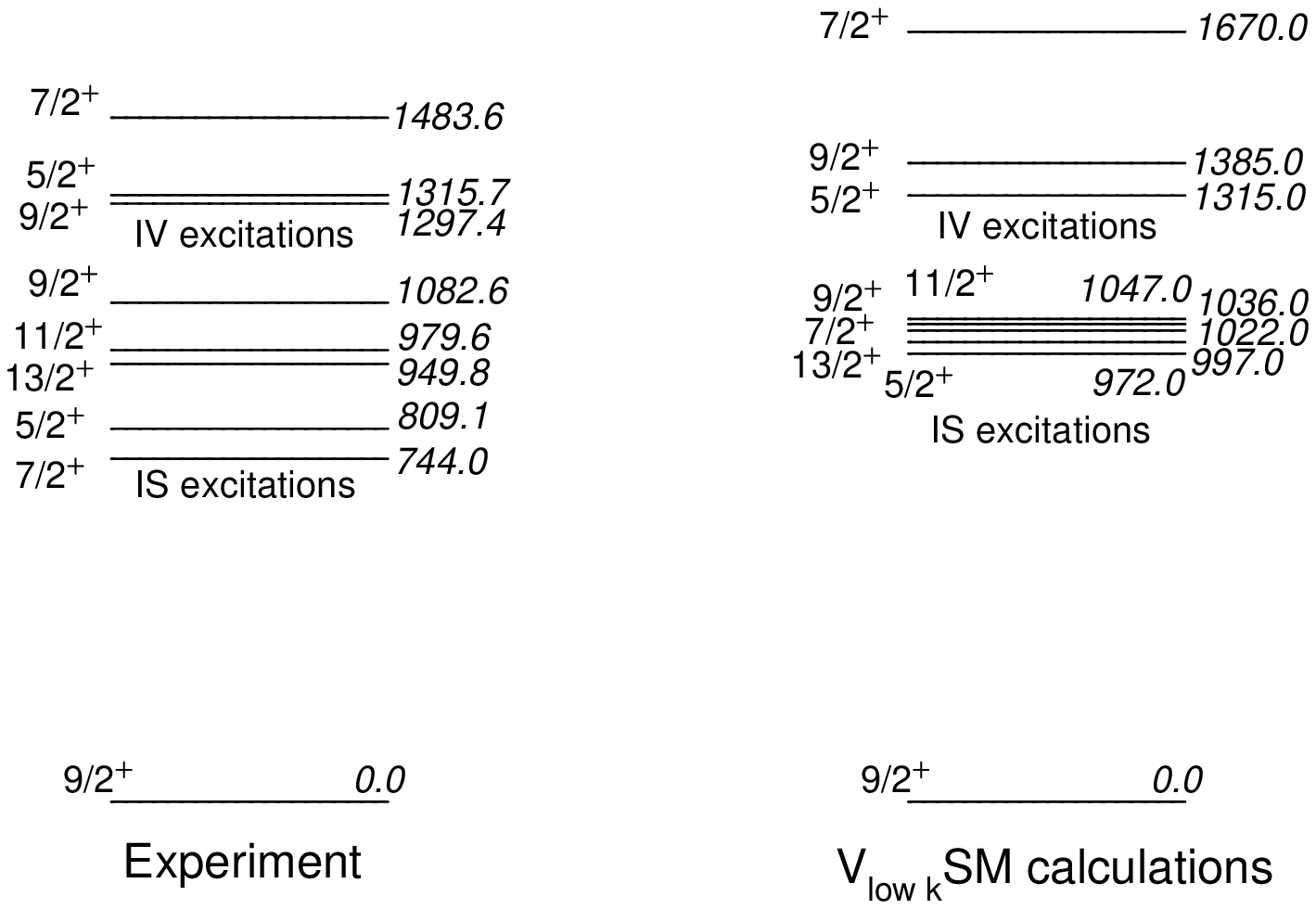}
\caption{$^{93}$Nb partial level scheme showing first-order $IS$ and $IV$ positive-parity 
excitations (left) as compared with {\vlk} $SM$ calculations (right).}
\label{fig:thvsexp}       
\end{center}
\end{figure}

To test this interaction, we have compared the
calculated proton-proton ({\it pp}) and neutron-neutron ({\it nn})
spectra with the experimentally observed spectra for the $\rm
^{90}Sr$ and $\rm ^{90}Zr$ nuclei up to excitation energies of
$\sim$ 3 MeV \cite{holtprep}. Here, we see there is
generally fair agreement between the calculated and experimental
levels, noting that the {\vlk} calculation gave rather similar
results to the those obtained in Ref.~\cite{3}, in which the
surface delta interaction, with tuned parameters, was used as the residual interaction. 
Particularly, Figure~\ref{fig:thvsexp} shows a partial 
level scheme with first-order $IS$ and $IV$ positive-parity 
excitations identified in this work compared with {\vlk} $SM$ calculations.
 The calculations were carried out using the OXBASH $SM$ code
\cite{oxbash} with a model space file and {\vlk} interaction file
specifically generated, using the above single-particle energies, for use
with this code.

For comparison with experimental work, we explicitly calculate the
transition rates, defined as:
\begin{equation}
B\left(M1:J_i\rightarrow J_f \right)=
\frac{\left| \left\langle J_f \right\| M1 \left\|J_i\right\rangle\right|^2}
{2J_i+1}.
\end{equation}
In these calculations we have kept the bare orbital $g$-factors, $g^l_{\pi}=1\mu_N$ and 
$g^l_{\nu}=0$, while we have used empirical values for the  spin $g$-factors, 
$g^s_{\pi}=3.18\mu_N$ and $g^s_{\nu}=-2.18\mu_N$.
We note that these values for the spin $g$ factors are not fit to
any experimental data.
Similarly, the $E2$ transition operator is given by
\begin{equation}
T\left(E2\right)=
e_{\pi}\sum^Z_{i=1}r^2_iY^{(2)}_{\mu}\left( \hat{r}_i\right)
+
e_{\nu}\sum^N_{i=1}r^2_iY^{(2)}_{\mu}\left( \hat{r}_i\right),
\end{equation}
where $e_{\pi}$ and $e_{\nu}$ are the proton and neutron effective
charges. Here, we use the same effective charges as in Ref.~\cite{orce}; that is, 
$e_{\pi}=1.85e$ and $e_{\nu}=1.30e$.

\subsection{Positive-Parity States}

In even-even nuclei the signature of large magnetic
dipole transition strength from a proposed isovector excitation  to the  isoscalar
state is typically sufficient for identification of $MS$ states.  Due to the presence of an unpaired
nucleon in an odd-mass nucleus, however, further theoretical evidence is needed
to confidently identify a $MS$ state. Here, in addition to the magnetic dipole transition
rates, we  decompose the relevant transitions into their spin and orbital
components to rule out  large $M1$ strengths due to  spin-flip transitions.  The calculations
are presented in Table \ref{tab:nb93m1}, where we have only included the results for
states where a clearly analogous state was experimentally present.

\subsubsection{First-order $IS$ excitations}
The lowest-lying quintet \{5/2$^+_1$, 7/2$^+_1$, 9/2$^+_2$, 11/2$^+_2$, 13/2$^+_2$\} can clearly be
identified as the coupling of the
g$_{9/2}$ proton to the 2$^+_1$  state in $^{92}$Zr.  While the experiments were not
sensitive to the ground-state decay rates of these states, the $SM$ calculations
reveal all to have strongly collective $B(E2)$ transition rates to the 9/2$^+$ ground state, with
predominantly isoscalar character.

\begin{table}
\centering \caption{Spin and orbital contributions to the large $M1$
transitions observed in the SM calculations.  Values are given in units
of $\mu_N^2$.}
\label{tab:nb93m1}
\vskip.1in
\begin{tabular}{clccc}
\hline\hline
\\
$J_i$&                $\rightarrow J_f$  & Spin $B(M1)$ & Orbital $B(M1)$ \\ [.1in]
\hline
\\
 5/2$^+_{2}  $&$ \rightarrow 7/2^+_{1} $ &  0.1434  & 0.1280  \\[.05in]
 7/2$^+_{3}  $&$ \rightarrow 9/2^+_{1} $ &  0.1248  & 0.09257 \\ [.05in]
              &$ \rightarrow 5/2^+_{1} $ &  0.05369 & 0.04515 \\ [.05in]
 9/2$^+_{3}  $&$ \rightarrow 7/2^+_{1} $ &  0.05797 & 0.06953 \\[.05in]
11/2$^+_{2}  $&$ \rightarrow 13/2^+_{1}$ &  0.03941 & 0.03586 \\ [.1in]
\hline\hline
\end{tabular}
\end{table}

\subsubsection{First-order $IV$ excitations}

We start with the 1279.4 keV 9/2$^+$ state, which exhibits a strong $M1$ transition to the
symmetric one-phonon 7/2$^+_1$ state and a weakly collective $E2$ transition to the 9/2$^+$ ground
state experimentally.  In the calculations, we find a qualitatively similar decay pattern,
though the rates are slightly underpredicted in the SM calculations.  In Table \ref{tab:nb93m1}, we see
that this magnetic dipole transition is almost equally composed of spin and orbital parts, as
is typically seen in such transitions \cite{orce}, leading to a confident assignment as a
MS excitation.

The 1483 keV 7/2$^+$ state exhibits a smiliar experimental decay pattern, with strong $M1$ transitions
to both the 9/2$^+_1$ and 5/2$^+_1$ states (though only an upper bound is determined for the transition
to the 9/2$^+_1$ state) and a weakly $E2$ transition to the ground state.  In the SM calculations
the large $M1$ transitions are again in qualitative agreement with the experimental measurements.
In the calculations the transition to the ground state, however, is significantly more collective than
the experimental value, but not inconsistent with its proposed identification as a MS excitation.  To confirm
this, we again turn to Table \ref{tab:nb93m1}, where we see that for both of these $M1$ transitions
there is significant orbital character, indicating a MS state.

From an inspection of $M1$ transition strength, it appears that the 1315 keV 5/2$^+$ state would be
a reasonable $IV$ excitation candidate with a strong transition to the 7/2$^+_1$ state.  This is coupled
with a collective $E2$ transition to the same state, indicative of an $IS$
transition.  This is perhaps due to mixing between this and the 1665.6-keV  5/2$^+$ state,  which also 
exhibits a weaker but sizable $M1$ transition to the 7/2$^+_1$ state as well as a collective
$E2$ transition to the same state.  The $SM$ presents a picture consistent with this assessment,
predicting an additional magnetic dipole transition to the 5/2$^+_1$ state not seen experimentally.
The MS character of these large $M1$ transitions can be confirmed in Table \ref{tab:nb93m1}, where
we see a sizable orbital component of the transitions.

Finally, a case can be made for proposing the 1603-keV 11/2$^+$ state as a member of the $IV$ quintet due
to the experimentally observed $M1$ transition to the 13/2$^+_1$ one-phonon symmetric state.  This is borne
out in the calculations, where we see the expected equal distribution of spin and orbital $M1$ transition
strength.  Again,
some mixing with higher $IS$ excitations is likely manifested in the collective $E2$ transition
to the one-phonon symmetric states seen experimentally, which are also apparent in the $SM$.

It would seem that a suitable candidate for the 13/2$^+$ $IV$ excitation is the 1686 keV state, with a
large $M1$ to both the 11/2$^+_1$ and 13/2$^+_1$ states observed experimentally.  This, however, is unfortunately
not apparent in the SM calculations as the $M1$ transitions predicted are too weak to support this identification.

\section{Conclusion}

In this work, we have investigated $IV$ and $IS$ excitations in the positive-
and negative-parity structures of $^{93}$Nb. Identifications are based on 
from $M1$ and $E2$ transition strengths, spin and parity assignments, and shell
model calculations. These findings support the  weak-coupling picture of
fermions and bosons in both $\pi2p_{1/2}^{-1}$$\otimes$$^{94}$Mo and
$\pi 1g_{9/2}$$\otimes$$^{92}$Zr configurations. The marked
separation of the positive- and negative-parity structures in
$^{93}$Nb facilitates the comprehension of such a relevant
interaction, specially in the simpler scenario for the negative-parity states. 
Overall, larger $B(M1)$ values are observed in this nucleus as compared with 
the even-A neighbors as the result of additional spin-flip effects.  
Levels  assigned as $IV$ excitations lie at lower energies than that observed for
the 2$_{1,MS}^+$ state in $^{92}$Zr (1.847 MeV).  Interestingly, similar transition
strengths of about 30 W.u. are found from the levels  characterised
as $IS$ excitations.

This work was supported by the U.S. National Science Foundation under Grant No. PHY-0652415, 
the Deutsche Forschungsgemeinschaft, Grant No. Jo 391/3-1 and by the U.S. Department of Energy
under DE-FC02-07ER41457 (UNEDF SciDAC Collaboration). TRIUMF receives federal funding via a contribution agreement 
through the National Research Council of Canada.


\end{document}